\documentclass[12pt,preprint]{aastex}

\begin{document}

\title{Filling in the Gaps in the 4.85~GHz Sky}

\author{Stephen E.\ Healey\altaffilmark{1,6}, Lars Fuhrmann\altaffilmark{2}, Gregory B.\ Taylor\altaffilmark{3,5},\\
Roger W. Romani\altaffilmark{1}, Anthony C.\ S.\ Readhead\altaffilmark{4}}

\altaffiltext{1}{Department of Physics/KIPAC, Stanford University, Stanford, CA 94305, USA}
\altaffiltext{2}{Max-Planck-Institut f\"ur Radioastronomie, Bonn, Germany}
\altaffiltext{3}{Department of Physics and Astronomy, University of New Mexico, Albuquerque, NM 87131, USA}
\altaffiltext{4}{Department of Astronomy, California Institute of Technology, Pasadena, CA 91125, USA}
\altaffiltext{5}{G.~B.~Taylor is also an Adjunct Astronomer at the National Radio Astronomy Observatory.}
\altaffiltext{6}{Email: {\tt sehealey@astro.stanford.edu}}

\begin{abstract}
We describe a 4.85~GHz survey of bright, flat-spectrum radio sources conducted with the Effelsberg 100~m
telescope in an attempt to improve the completeness of existing surveys, such as CRATES.  We report the
results of these observations and of follow-up 8.4~GHz observations with the VLA of a subset of the sample.
We comment on the connection to the {\it WMAP} point source catalog and on the survey's effectiveness at
supplementing the CRATES sky coverage.
\end{abstract}


\section{Introduction}

\ Extensive effort has been made to survey the sky at 4.85~GHz with single-dish telescopes.  The
Parkes-MIT-NRAO survey (PMN; \citealt{pmn}), conducted with the 64~m telescope at Parkes, covers the
region $-87\arcdeg < \delta < +10\arcdeg$ to a flux density limit of $\sim$30~mJy.  The Green Bank
6~cm survey (GB6; \citealt{gb6}), conducted with the former 91~m telescope at Green Bank, covers the
region $0\arcdeg < \delta < +75\arcdeg$ to a limit of $\sim$18~mJy.  The Fifth 5~GHz Strong
Source Survey (S5; \citealt{s5}), conducted with the Effelsberg 100~m telescope, covers the
region $+70\arcdeg < \delta < +90\arcdeg$ to a limit of $\sim$250~mJy.  Together, they constitute
a (non-uniform) catalog of almost 125,000 distinct radio sources over nearly the entire sky.

\ Data from these surveys were subsequently used to select targets for more specialized study.  The
Cosmic Lens All-Sky Survey (CLASS; \citealt{class,class2}) identified sources that were bright in GB6 and
had flat spectra between the 1.4~GHz NRAO VLA Sky Survey (NVSS; \citealt{nvss}) and GB6; these were
then observed at 8.4~GHz with the VLA to look for gravitationally lensed compact radio sources.
Although the objective of CLASS was to find lenses, \citet{srm3} demonstrated its usefulness in
identifying $\gamma$-ray blazars, which correlate strongly with bright, flat-spectrum radio sources.
The Combined Radio All-sky Targeted Eight~GHz Survey (CRATES; \citealt{crates}) extended the CLASS
procedure to the entire sky with the explicit goal of identifying blazar candidates.  It used the
full complement of 4.85~GHz sources to identify over 11,000 bright sources with flat spectra, which
were followed up with 8.4~GHz interferometry.

\ Much of the utility of the CRATES sample, especially with regard to identifying blazar candidates,
is in its breadth of coverage and its uniformity.  \citeauthor{srm3}, for example, developed a
figure of merit for associating radio sources with $\gamma$-ray detections.  The ability to do so
in a statistically meaningful way rested on the uniform selection criteria and large area coverage
of CLASS.  Similarly, the Candidate Gamma-Ray Blazar Survey (CGRaBS; \citealt{cgrabs}) is a
catalog of the $\sim$15\% of CRATES sources that are most similar to the Third EGRET Catalogue
(3EG; \citealt{3eg}) blazars as quantified by a radio/X-ray figure of merit.  Without the large
number of consistently selected sources provided by CRATES, a catalog with the statistical power
of CGRaBS would be impossible to compile.  Improving the sample's coverage is further motivated by
comparison with other all-sky surveys, such as the {\it Wilkinson Microwave Anisotropy Probe}
(\hspace{-0.15em}{\it WMAP}\hspace{0.1em}) five-year point source catalog \citep{wmap5}.  This
shallow but uniform high-frequency radio survey of the entire sky is composed mostly of bright,
flat-spectrum sources; thus, the identification of low-frequency counterparts depends on the
availability of all-sky low-frequency data.

\ It is therefore of some importance to note that the sky coverage of the large 4.85~GHz
surveys---S5, GB6, and PMN---is not perfect.  In the north polar cap (an area of $\sim$700 square
degrees), the S5 survey is considerably shallower than GB6 and PMN, as noted above.  There are also
two ``holes'' in the PMN survey just south of the equator (an area of $\sim$300 square degrees) in
which no data are available at all.  These holes are already known to contain $\gamma$-ray blazars
in 3EG and in the list of bright AGNs \citep{latagn} detected in the first three months of data from
the Large Area Telescope (LAT) on board the {\it Fermi Gamma-ray Space Telescope}.  Likewise, it is
expected that $\gamma$-ray blazars not associated with S5 sources will also be detected by the LAT
in the far north.  The real solution to this problem is a full 4.85~GHz survey of the entire
northern cap and PMN hole regions down to the CRATES flux density limit ($\sim$65~mJy) or better;
such a campaign would require approximately a week of observing time.
However, in the interest of time and logistics, we have conducted a targeted 4.85~GHz survey,
requiring only $\sim$24 hours of observing time, of selected sources in these regions with the
Effelsberg 100~m telescope as an initial step toward filling in the gaps in the CRATES all-sky
coverage.

\section{Sample Selection}

\subsection{Sky Regions}

\ Our work concerns two distinct regions on the sky.  The ``north polar cap'' is the region
$+75\arcdeg < \delta < +90\arcdeg$, spanning all right ascensions.  The ``PMN holes'' are two
nearly parallelogram-shaped regions just south of the equator.  Their vertices, given as
$(\alpha,\delta)$ in decimal degrees (J2000 coordinates), are approximately as follows:

\noindent PMN hole 1: $(175\arcdeg,0\arcdeg)$; $(183\arcdeg,0\arcdeg)$; $(197\arcdeg,-9.5\arcdeg)$; $(188\arcdeg,-9.5\arcdeg)$.
\\ PMN hole 2: $(195\arcdeg,0\arcdeg)$; $(217\arcdeg,0\arcdeg)$; $(235\arcdeg,-9.5\arcdeg)$; $(210\arcdeg,-9.5\arcdeg)$.

\subsection{Choosing Sources}

\ Our goal was to observe flat-spectrum targets that were likely to be bright at 4.85~GHz.  In
order to select such sources, we drew from archival radio surveys at lower frequencies.  NVSS
provides 1.4~GHz coverage of the entire $\delta > -40\arcdeg$ sky, but to identify sources with
flat spectra, data at a second frequency are required.  In the north polar cap, 325~MHz data are
available from the Westerbork Northern Sky Survey (WENSS; \citealt{wenss}) while in the PMN
holes, 365~MHz data are available from the Texas Survey \citep{texas}.  We identified NVSS
sources with counterparts in WENSS/Texas and computed the spectral index of each source (assuming
a power law spectrum $S \propto \nu^\alpha$).  From this, we calculated a predicted 4.85~GHz flux
density ($S_\mathrm{pred}$) for each source.  Due to logistical considerations at Effelsberg, the
final selection criteria were slightly different for the two observing regions.  In the north
polar cap, we selected sources with $S_\mathrm{pred} > 65$~mJy and $\alpha > -0.6$ that were not
already included in the S5 survey (221 sources).  It is worth noting that S5 itself only contains
236 sources, so our observations nearly double the coverage in this region.  In the equatorial
holes, we selected sources with $S_\mathrm{pred} > 75$~mJy and $\alpha > -1.0$ (174 sources).

The top panel of Figure~1 shows a sky map of the sources in the CRATES catalog.  The bottom panel
shows the the additional sources selected for observation at Effelsberg.

\section{Observations and Data Analysis}
\ The flux density measurements of the selected sources were performed with the Effelsberg 100~m
telescope of the Max-Planck-Institut f\"ur Radioastronomie (MPIfR) during two observing runs in
June and July 2008 with a total observing time of $\sim$24 hrs.  We used the 4.85~GHz multi-horn
receiver permanently mounted at the secondary focus for sensitive continuum measurements.  The
4.85~GHz system is a double-horn heterodyne receiver that allows the subtraction of the
off-source atmospheric contribution from the on-source astronomical signal.  Both circular
polarizations are fed into a broad-band polarimeter, measuring the total power (Stokes I).

\ The target sources are sufficiently bright in the observed frequency band to allow flux density
measurements using cross-scans in azimuth and elevation.  In contrast to the on-off observing
technique, this enables an instantaneous control of the telescope pointing as well as the
identification of possible problems with confused or extended sources.  Sources with
$S_\mathrm{pred} \ge 100$~mJy were observed with four cross-scans (two in each direction), and
sources with $S_\mathrm{pred} < 100$~mJy were observed with six cross-scans (three in each
direction).

\ In order to minimize slewing time, we observed sources along stripes parallel to the right
ascension axis and about $3\arcdeg$ wide in declination.  In the case of particularly weak
sources (or obviously bad scans), sources were subsequently re-observed to improve S/N.  We
observed primary flux density calibrators (e.g., 3C~286, 3C~295 and NGC~7027) frequently to
adjust the focus of the telescope and to link the measured flux densities to the absolute flux
density scale \citep{baars,ott}.

\ The first steps in the data reduction included baseline subtraction and fitting of a Gaussian profile 
to each individual sub-scan in both slewing directions: the peak of the Gaussian measures the 
antenna temperature on the source, the HPBW ($\sim$$144\arcsec$), and the positional displacement 
of the peak (i.e., the residual pointing error of the telescope, typically $<$$10\arcsec$).  We flagged
bad scans or sub-scans and excluded them from further analysis.  For sources showing confusion problems
(i.e., a secondary source in the off-horn position, causing severe distortion), we reduced the data
without the software beam-switch.  After correcting the measured amplitudes for residual positional
offsets using the approximately Gaussian shape of the telescope beam, we averaged them over both
slewing directions.  We performed an opacity correction using the $T_\mathrm{sys}$ measurements
obtained for each sub-scan and a standard correction for the systematic elevation-dependent telescope
gain.  The frequent primary calibrator measurements allowed us to convert the measured antenna
temperatures for each target into absolute flux densities. 

\section{Results}
In total, we observed 395 sources across both sky regions.  For 368 (93\%) of the sources, we
obtained precise flux densities with a mean fractional uncertainty of 2.3\%.  The individual measurement
uncertainties result mostly from statistical errors in the reduction process (including errors from the
Gaussian fit) and a contribution from the scatter seen in the primary calibrator measurements.  By
independently studying a large sample of weak sources at 4.85~GHz, \citet{angel} showed that for flux
densities $\la$100~mJy, the dominant sources of uncertainty are thermal noise, confusion, and
tropospheric instabilities.  Together, they result in an uncertainty of about 1.2~mJy, which we also
include in our error estimates.  The 4.85~GHz flux densities and uncertainties are given in
Table~1 (for the sources in the north polar cap) and Table~2 (for the sources in the PMN holes).  The
NVSS positions were sufficient given the pointing accuracy of the Effelsberg 100~m telescope, and they
are more accurate than can be measured by the 100~m telescope, so we report the NVSS positions in the
tables.

\section{VLA Follow-up}
We conducted a follow-up campaign on 109 sources (72 in the north polar cap and 37 in the PMN holes),
with the VLA in the A configuration, using two 50~MHz bands at 8.44~GHz.  We targeted sources with
$S_\mathrm{4.85\;GHz} \ge 65$~mJy from our Effelsberg observations and $\alpha > -0.6$ between 1.4~GHz
(from NVSS) and 4.85~GHz.  The observations
took place on 2008 November 1 as part of program AH0976, and the on-source dwell time was 60~s.
Standard AIPS calibration was performed, followed by Difmap imaging and Gaussian component fitting.
We used 3C~286 as a flux density calibrator, and the typical radiometric error is approximately 3\%.
With interferometric X-band measurements in hand, these sources are now completely on par with those
in CRATES.

The results of these observations are shown in Tables~1 and 2.  We report the position and integrated
flux density of each source.  For sources with multiple components, we report the position and
integrated flux density of each individual component.  The typical positional accuracy is
$\la$$0.06\arcsec$, which is consistent with the positional uncertainties of the calibrators and the
expected positional accuracy under good conditions.  Four sources (J0033+7829, J0202+8115,
J1445$-$0329, and J2329+7808) have components that were resolved by the VLA; these are less likely to
be the compact cores of blazars and are indicated in the tables.

\section{{\it WMAP} Point Sources}

\ The {\it WMAP} five-year data release includes a list of 390 point sources detected in the sky
maps of at least one of the {\it WMAP} frequency bands (K~=~23~GHz, Ka~=~33~GHz, Q~=~41~GHz,
V~=~61~GHz, W~=~94~GHz).  This provides an all-sky high-frequency radio
catalog of the sky against which to check our selection of bright, flat-spectrum sources.
\citeauthor{crates}\ showed that the vast majority ($\sim$88\%) of sources in the {\it WMAP}
three-year point source catalog \citep{wmap3} had counterparts in CRATES.  Those that did not were
dominated by sources that were very bright but had steep spectra.  However, some of the sources
without counterparts were bright, flat-spectrum sources that were simply located in regions not
covered by CRATES (e.g., the Galactic plane and the PMN holes).  At the {\it WMAP} areal density
($9.5 \times 10^{-3}$~{deg$^{-2}$), we expect the PMN holes to contain $\sim$3 {\it WMAP} sources, and
indeed, two sources (WMAP~J1408$-$0749 and WMAP~J1512$-$0904) are located there and another
(WMAP~J1510$-$0546) is right on the edge of the PMN coverage.  Our sample selection included these
three sources (i.e., they were not specially targeted), and our 4.85~GHz data confirm that they are
indeed bright, flat-spectrum sources like those in the CRATES sample.

\section{Discussion}

Figure~2 shows a comparison of the observed 4.85~GHz flux densities to the flux densities predicted as
described in \S 2.2.  As the distribution shows, our simple power-law extrapolation is not a very good
indicator of the true flux density; it overpredicts the observed value for $\sim$80\% of the targets,
and $\langle S_\mathrm{obs}/S_\mathrm{pred} \rangle$ = 0.75.  As a result, many of the observed sources
do not end up satisfying the nominal CRATES cuts on flux density ($S_\mathrm{4.8\;GHz} \ge 65$~mJy)
and/or spectral index ($\alpha > -0.5$).  Indeed, by requiring a bright detection at $\sim$0.35~GHz, we
seem to have selected a significant number of gigahertz-peaked spectrum (GPS) sources rather than strong
flat-spectrum 4.85~GHz sources.  Figure~3 shows a scatter plot of the spectral index between
1.4~GHz and 4.85~GHz (``$\alpha_\mathrm{NVSS/Eff}$'') vs.\ the spectral index between $\sim$0.35~GHz and
1.4~GHz (``$\alpha_\mathrm{low/NVSS}$'').  The shaded region represents sources with spectra that rise from
$\sim$0.35~GHz to 1.4~GHz but fall from 1.4~GHz to 4.85~GHz (i.e., GPS sources); our catalog contains 75
such sources.  Note also that most sources ($>$80\%) fall below the dashed line, indicating spectra that
are steeper above 1.4~GHz than below it.  An alternative approach to sample selection, such as simply
targeting the brightest NVSS sources, may have yielded a higher fraction of CRATES-like sources.

Nevertheless, the results of this survey are of value.  Of the $\sim$160 CRATES-like sources
sources missing in the north polar cap, we have identified 57, bringing the coverage there from
$\sim$33\% to $\sim$57\%.  Of the $\sim$100 missing sources in the PMN holes, we have identified 24,
bringing the coverage there from 0\% to $\sim$24\%.  The total coverage of the $|b| > 10\arcdeg$ sky
is increased from 97.6\% to 98.3\%.  This catalog of strong, compact sources with subarcsecond
astrometry is also of great use for calibrating radio telescope arrays.  Additionally, our 4.85~GHz
survey, along with the 8.4~GHz follow-up measurements, brings the fraction of high-latitude {\it WMAP}
point sources with CRATES-like counterparts to 84\%.  Our observations will be even more important in
the {\it Planck} era; the increase in sensitivity ($\sim$30$\times$) over {\it WMAP} will yield
thousands of point sources over the entire sky, and the identification of counterparts will rely
heavily on catalogs of bright, flat-spectrum radio sources.

\acknowledgements
Data in this paper are based on observations with the 100~m telescope of the MPIfR
(Max-Planck-Institut f\"ur Radioastronomie) at Effelsberg.  The National Radio Astronomy Observatory
is operated by Associated Universities, Inc., under cooperative agreement with the National Science
Foundation.  S.~E.~H.\ was supported by SLAC under DOE contract DE-AC03-76SF00515.

\begin{deluxetable}{cccrrccr}
\setlength{\tabcolsep}{0.1in}
\tabletypesize{\scriptsize}
\tablewidth{0pt}
\tablecaption{Sources in the north polar cap}
\tablehead{
  \colhead{Name}&
  \multicolumn{2}{c}{NVSS position}&
  \colhead{$S_{4.85}$}&
  \colhead{$\sigma_{4.85}$}&
  \multicolumn{2}{c}{VLA 8.4 GHz position}&
  \colhead{$S_{8.4}$\tablenotemark{a}}\\
  
  &
  \multicolumn{2}{c}{(J2000)}&
  \colhead{(mJy)}&
  \colhead{(mJy)}&
  \multicolumn{2}{c}{(J2000)}&
  \colhead{(mJy)}
}

\startdata
J0010$+$7614&00:10:17.83&$+$76:14:19.8&25.4&1.2\\
J0022$+$7501&00:22:02.13&$+$75:01:22.5&34.0&1.3\\
J0029$+$8031&00:29:27.15&$+$80:31:31.6&80.0&1.6\\
J0032$+$8750&00:32:41.68&$+$87:50:43.7&57.1&1.4\\
J0033$+$7829&00:33:08.23&$+$78:29:00.8&67.4&1.5&00:33:08.211&$+$78:29:00.12&48.5\tablenotemark{b}\\
J0045$+$8810&00:45:02.55&$+$88:10:17.9&66.6&1.5\\
J0054$+$7558&00:54:30.73&$+$75:58:08.0&66.6&1.5\\
J0115$+$8109&01:15:42.01&$+$81:09:54.2&52.2&1.4\\
J0128$+$7928&01:28:08.93&$+$79:28:46.1&113.3&1.9&01:28:08.870&$+$79:28:46.14&93.0\\
J0138$+$8611&01:38:29.65&$+$86:11:40.1&68.0&1.5&01:38:29.637&$+$86:11:40.90&38.7\\
J0143$+$8544&01:43:35.52&$+$85:44:22.7&82.0&1.6&01:43:35.551&$+$85:44:23.50&68.4\\
J0144$+$7958&01:44:37.08&$+$79:58:39.5&19.6&1.2\\
J0151$+$8303&01:51:18.31&$+$83:03:20.2&51.6&1.4\\
J0157$+$7552&01:57:11.56&$+$75:52:28.8&102.6&1.8\\
J0202$+$8115&02:02:05.42&$+$81:15:45.0&163.7&2.4&02:02:05.739&$+$81:15:45.28&47.6\tablenotemark{b}\\
J0211$+$8449&02:11:41.58&$+$84:49:47.7&78.4&1.6&02:11:41.338&$+$84:49:47.92&55.5\\
J0215$+$7554&02:15:17.69&$+$75:54:53.6&96.6&1.7&02:15:17.902&$+$75:54:53.01&68.9\\
J0249$+$8019&02:49:40.73&$+$80:19:25.3&38.6&1.3\\
J0249$+$8435&02:49:48.43&$+$84:35:56.4&126.7&2.0&02:49:48.331&$+$84:35:57.03&121.0\\
J0304$+$7930&03:04:56.81&$+$79:30:52.2&26.4&1.2\\
J0309$+$8601&03:09:42.29&$+$86:01:39.7&27.0&1.3\\
J0316$+$7720&03:16:32.19&$+$77:20:59.2&115.6&1.9&03:16:32.143&$+$77:20:58.40&133.7\\
J0317$+$7658&03:17:51.96&$+$76:58:33.9&83.2&1.6&03:17:54.072&$+$76:58:38.06&13.4\\
J0320$+$8711&03:20:21.87&$+$87:11:18.6&50.6&1.4\\
J0324$+$7849&03:24:14.60&$+$78:49:10.9&106.1&1.8&03:24:14.531&$+$78:49:11.65&82.6\\
J0325$+$8517&03:25:11.23&$+$85:17:40.0&82.9&1.6&03:25:11.125&$+$85:17:39.45&42.7\\
&&&&&03:25:13.732&$+$85:17:43.70&4.4\\
J0410$+$8208&04:10:39.50&$+$82:08:23.3&115.3&1.9\\
J0411$+$8509&04:11:52.15&$+$85:09:43.9&52.5&1.4\\
J0428$+$7528&04:28:18.59&$+$75:28:00.3&18.8&1.2\\
J0428$+$8314&04:28:21.90&$+$83:14:55.3&48.7&1.4\\
J0440$+$7755&04:40:18.91&$+$77:55:58.5&56.3&1.4\\
J0442$+$8742&04:42:25.51&$+$87:42:57.4&21.3&1.2\\
J0449$+$8233&04:49:00.25&$+$82:33:18.8&197.8&2.8&04:49:00.184&$+$82:33:19.45&150.7\\
J0501$+$8505&05:01:54.66&$+$85:05:19.9&15.3&1.2\\
J0517$+$8108&05:17:57.65&$+$81:08:51.9&13.3&1.2\\
J0524$+$7627&05:24:22.17&$+$76:27:35.4&58.6&1.4\\
J0532$+$7623&05:32:29.80&$+$76:23:28.3&56.8&1.4\\
J0540$+$7702&05:40:52.10&$+$77:02:01.5&46.8&1.3\\
J0546$+$7943&05:46:48.52&$+$79:43:25.2&31.1&1.3\\
\enddata

\tablecomments{Only the first page of Table 1 is shown here.  The table is
published in its entirety in the electronic edition.}

\tablenotetext{a}{The typical error in the 8.4 GHz flux density is approximately 3\%.}

\tablenotetext{b}{Component is resolved or partially resolved.}

\end{deluxetable}

\begin{deluxetable}{cccrrccr}
\setlength{\tabcolsep}{0.1in}
\tabletypesize{\scriptsize}
\tablewidth{0pt}
\tablecaption{Sources in the PMN holes}
\tablehead{
  \colhead{Name}&
  \multicolumn{2}{c}{NVSS position}&
  \colhead{$S_{4.85}$}&
  \colhead{$\sigma_{4.85}$}&
  \multicolumn{2}{c}{VLA 8.4 GHz position}&
  \colhead{$S_{8.4}$\tablenotemark{a}}\\
  
  &
  \multicolumn{2}{c}{(J2000)}&
  \colhead{(mJy)}&
  \colhead{(mJy)}&
  \multicolumn{2}{c}{(J2000)}&
  \colhead{(mJy)}
}

\startdata
J1144$-$0031&11:44:54.01&$-$00:31:36.6&215.8&3.1\\
J1146$-$0007&11:46:40.93&$-$00:07:33.7&42.6&1.3\\
J1146$-$0108&11:46:42.88&$-$01:08:05.1&121.8&2.0\\
J1148$-$0046&11:48:07.19&$-$00:46:45.3&165.5&2.5\\
J1150$-$0023&11:50:43.88&$-$00:23:54.4&1797.8&23.4\\
J1156$-$0212&11:56:38.19&$-$02:12:53.8&71.8&1.5\\
J1157$-$0124&11:57:11.03&$-$01:24:11.0&69.9&1.5\\
J1157$-$0312&11:57:50.08&$-$03:12:23.8&148.2&2.3\\
J1202$-$0240&12:02:32.28&$-$02:40:03.3&219.9&3.1\\
J1202$-$0336&12:02:51.43&$-$03:36:26.6&97.5&1.7\\
J1204$-$0029&12:04:05.69&$-$00:29:50.0&157.7&2.4&12:04:05.683&$-$00:29:53.55&50.7\\
J1204$-$0422&12:04:02.13&$-$04:22:44.0&962.7&12.6\\
J1207$-$0106&12:07:41.66&$-$01:06:37.4&153.5&2.3&12:07:41.679&$-$01:06:36.67&107.7\\
&&&&&12:07:42.244&$-$01:06:47.11&6.5\\
&&&&&12:07:41.589&$-$01:06:39.67&4.9\\
J1209$-$0257&12:09:06.88&$-$02:57:45.2&106.5&1.8\\
J1210$-$0136&12:10:31.37&$-$01:36:50.1&258.0&3.6\\
J1210$-$0341&12:10:18.86&$-$03:41:54.0&65.4&1.5\\
J1213$-$0159&12:13:14.00&$-$01:59:04.0&52.5&1.4\\
J1214$-$0306&12:14:50.51&$-$03:06:49.2&86.5&1.6&12:14:50.517&$-$03:06:49.55&65.0\\
J1214$-$0416&12:14:32.37&$-$04:16:03.4&79.6&1.6\\
J1214$-$0606&12:14:35.91&$-$06:06:58.9&86.0&1.6\\
J1215$-$0628&12:15:14.42&$-$06:28:03.6&308.0&4.2&12:15:14.394&$-$06:28:03.90&296.9\\
J1217$-$0337&12:17:55.30&$-$03:37:21.2&73.3&1.5\\
J1218$-$0631&12:18:36.18&$-$06:31:15.9&181.4&2.6\\
J1221$-$0241&12:21:23.90&$-$02:41:50.9&448.0&5.9&12:21:23.942&$-$02:41:49.60&210.8\\
&&&&&12:21:23.921&$-$02:41:49.84&31.7\\
&&&&&12:21:30.395&$-$02:41:32.99&23.0\\
&&&&&12:21:30.409&$-$02:41:32.95&18.7\\
J1222$-$0449&12:22:24.14&$-$04:49:33.0&62.3&1.4\\
J1226$-$0421&12:26:46.60&$-$04:21:18.6&104.2&1.8\\
J1226$-$0434&12:26:56.76&$-$04:34:22.2&116.1&1.9\\
J1227$-$0445&12:27:16.53&$-$04:45:32.8&39.5&1.3\\
J1228$-$0631&12:28:52.20&$-$06:31:46.3&191.7&2.8&12:28:52.176&$-$06:31:46.55&70.6\\
&&&&&12:28:52.192&$-$06:31:46.54&29.5\\
&&&&&12:28:52.209&$-$06:31:46.48&15.6\\
J1228$-$0838&12:28:20.43&$-$08:38:13.1&281.7&3.9\\
J1230$-$0510&12:30:29.00&$-$05:10:01.0&76.0&1.6\\
J1232$-$0717&12:32:52.31&$-$07:17:29.9&172.3&2.5\\
J1233$-$0613&12:33:31.27&$-$06:13:22.8&67.1&1.5\\
\enddata

\tablecomments{Only the first page of Table 2 is shown here.  The table is
published in its entirety in the electronic edition.}

\tablenotetext{a}{The typical error in the 8.4 GHz flux density is approximately 3\%.}

\tablenotetext{b}{Component is resolved or partially resolved.}

\end{deluxetable}

\begin{figure}[h]
\centering
\includegraphics[width=0.9\textwidth,trim=11mm 22mm 0mm 95mm,clip]{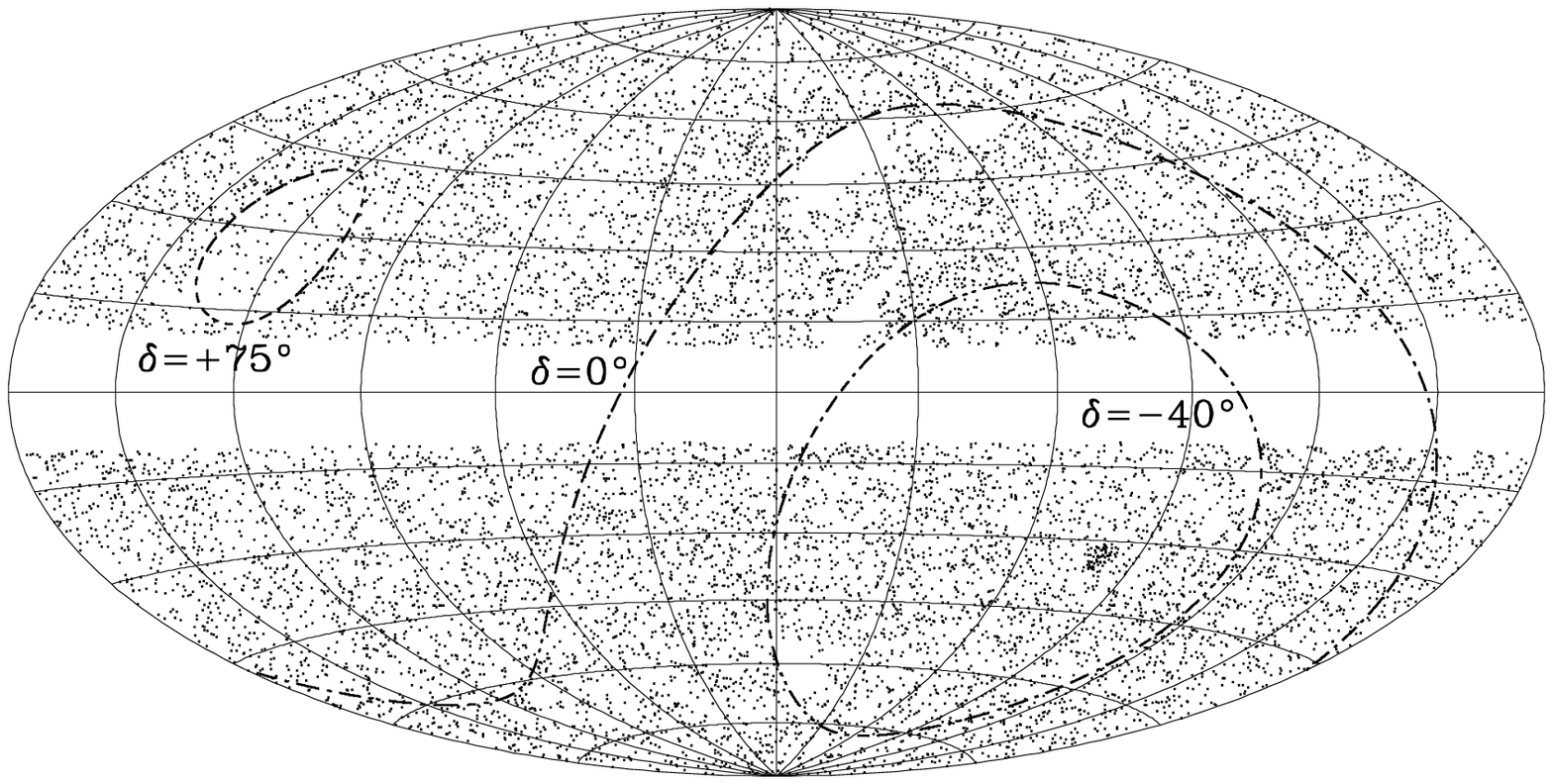}
\includegraphics[width=0.9\textwidth,trim=11mm 22mm 0mm 90mm,clip]{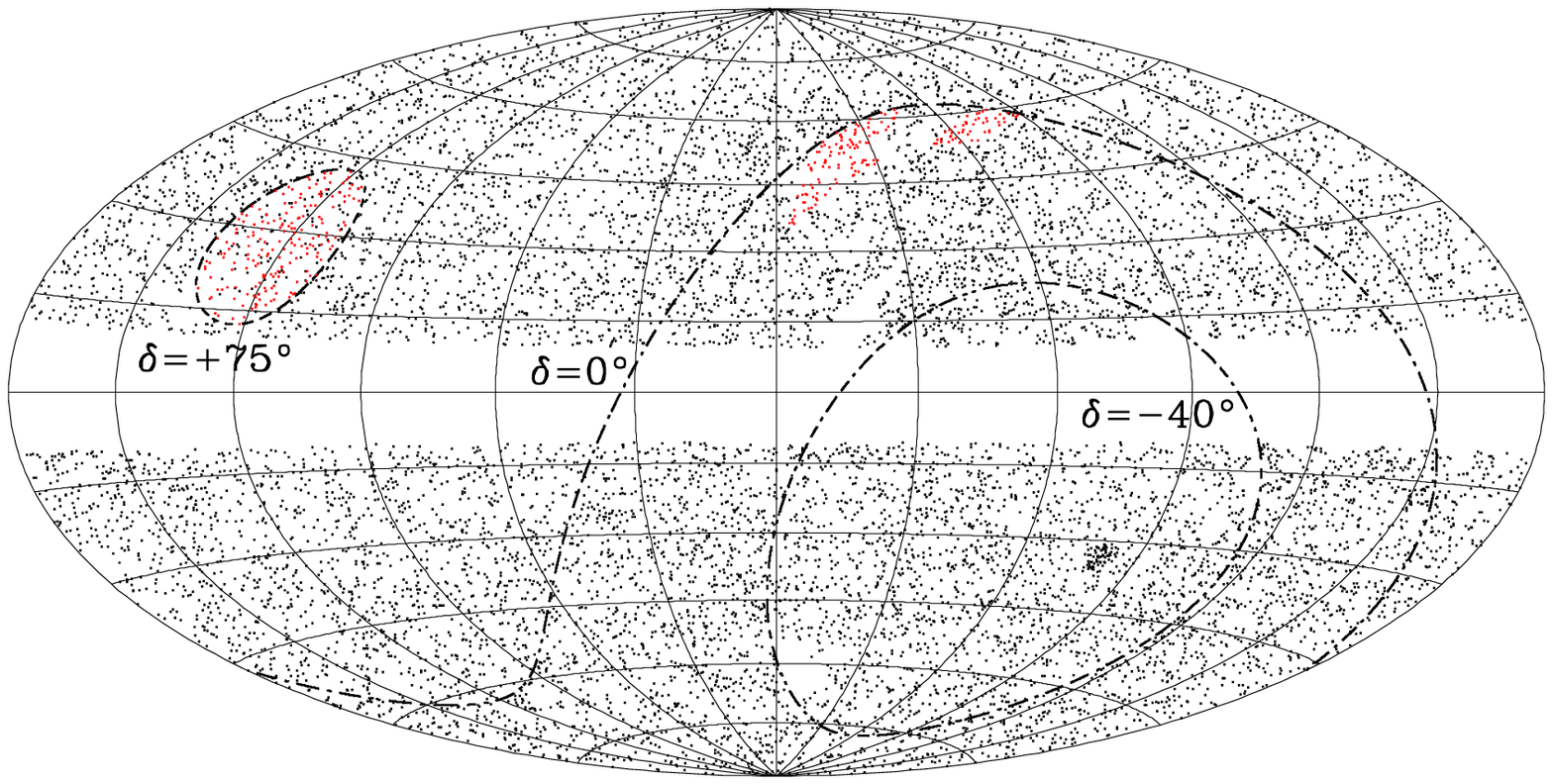}
\caption{{\it Top:} Aitoff equal-area projection of the CRATES catalog in Galactic coordinates $(l, b)$
with $l = 0\arcdeg$ at the center.
{\it Bottom:} The same sky map with the additional Effelsberg targets shown in red.}
\end{figure}

\begin{figure}[h]
\centering
\includegraphics[width=0.9\textwidth]{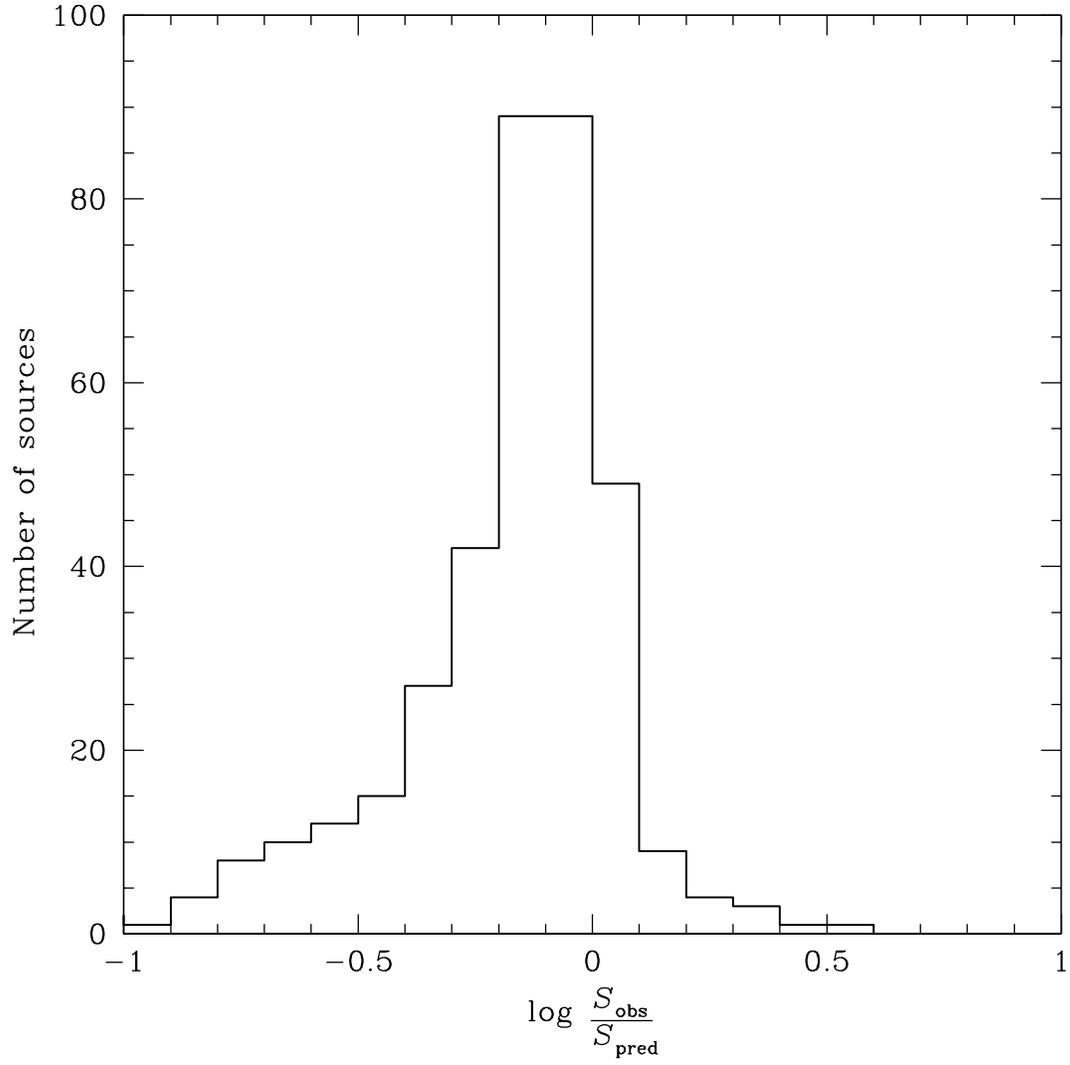}
\caption{Comparison of the observed values of the 4.85~GHz flux densities to the predicted values based
on power-law extrapolation from archival data at $\sim$0.35~GHz and 1.4~GHz.}
\end{figure}

\begin{figure}[h]
\centering
\includegraphics[width=0.9\textwidth]{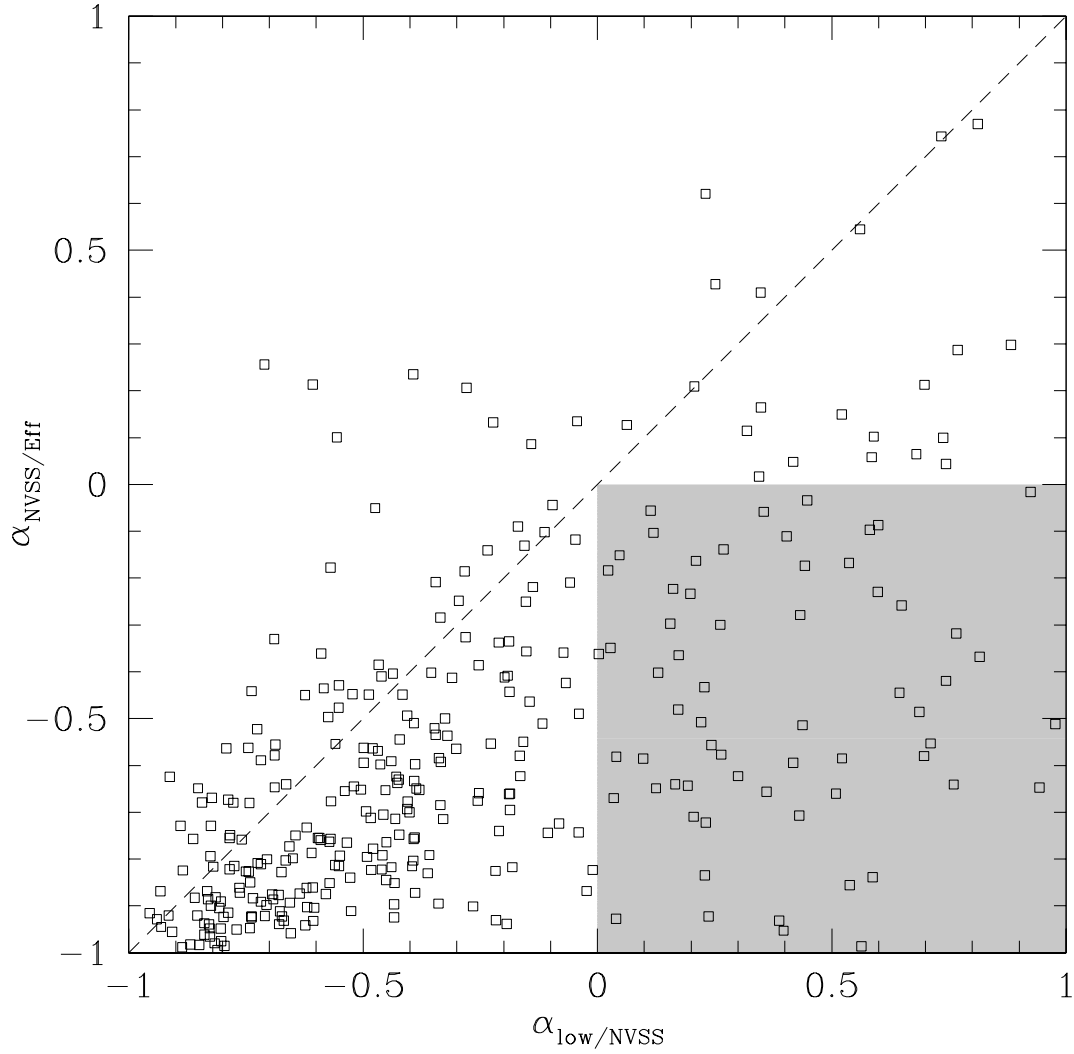}
\caption{Comparison of spectral indices from 1.4~GHz to 4.85~GHz (``$\alpha_\mathrm{NVSS/Eff}$'') to
spectral indices from $\sim$0.35~GHz to 1.4~GHz (``$\alpha_\mathrm{low/NVSS}$'').  Gigahertz-peaked
spectrum (GPS) sources lie in the shaded region.  Sources below the dashed line are steeper above
1.4~GHz than below it.}
\end{figure}

\end{document}